# Vortex-like Current States in Josephson Ballistic Point Contacts

A.N. Omelyanchouk, S.N. Shevchenko, and Yu.A. Kolesnichenko

*B.I. Verkin Institute for Low Temperature Physics and Engineering, 47 Lenin Ave., 61103, Kharkov, Ukraine*

*The stationary Josephson effect is studied theoretically in the situation when there is externally injected transport supercurrent which flows in the banks parallel to the contact interface. Coexistence of this supercurrent with the order parameter phase difference between the banks of the contact results in the appearance of quasiparticles in the weak link, which create the current, localized in the vicinity of the contact and counter-flowing to the transport supercurrent. We review the results of our previous study of weak links between current-carrying conventional and unconventional (d-wave) superconductors. And further we study the weak link in the form of double point contact between current-carrying superconductors. Current distribution patterns containing vortex-like states are obtained.*

## 1. INTRODUCTION: LOCAL AND NONLOCAL CURRENT STATES

The Josephson current in mesoscopic superconducting weak links is essentially nonlocal. It nonlocally depends on the spatial distribution of the phase $\chi(\mathbf{r})$ of the superconducting order parameter and is determined by the total phase difference $\phi$ across the weak link. This case is opposite to the homogeneous current state in which the supercurrent density $\mathbf{j}(\mathbf{r})$ locally depends on the superfluid velocity $\mathbf{v}_s = (\hbar/2m)\nabla\chi(\mathbf{r})$. A similar situation takes place in a normal metal ballistic point contact (a microconstriction of size much smaller than the electron mean free path). In the latter case the current density in the microconstriction is related to the total voltage difference $V$ on the contact, but not to the local value of the electric field $E(\mathbf{r})$; the total current through the contact $I$ equals $V/R_{Sh}$ ($R_{Sh}$ is the Sharvin resistance). The effects of nonlocality in Josephson microstructures

**A.N. Omelyanchouk, S.N. Shevchenko, Yu.A. Kolesnichenko**

were studied experimentally[1] and theoretically.[2–5] In particular, the effects of phase dragging and magnetic flux transfer in Josephson multiterminals were predicted[4] and studied.[5] The former effect is similar to the voltage dragging in normal metal mesoscopic multiterminals.[6]

The question about the interplay between nonlocal and local coherent superconducting states was raised recently.[7] It was considered, probably, as unique possibility for mixing of two current states of different nature, when the mesoscopic superconducting weak link is simultaneously subjected to the order parameter phase difference $\phi$ on the contact and to the tangential component of the superfluid velocity $\mathbf{v}_s$ in the banks. We have studied the Josephson junction between current-carrying superconductors (both conventional and unconventional) in the papers;[7–9] the results are reviewed in Sec. 2.

In the present paper we proceed with the nonlocal mixing of supercurrents in a system more complicated than the single-point contact, namely, in the double-point contact geometry with transport supercurrents in the banks (see Fig. 1c and comments in the next section). The single-point contact problem is a particular case of a double-point contact problem in the limit when the distance between the two orifices $2L$ tends to 0. When $L$ is small, this corresponds to a point contact with a defect. When $L$ is comparable with the size of an orifice $a$, then our consideration can give an idea of what happens in the case of the partition between banks with many orifices (which can be, for example, the punctures in the insulating layer between two superconductors). Also the double contact configuration can be considered as the implementation of mesoscopic dc SQUID (section 4.3). Josephson junctions between $d$-wave superconductors attract interest from the standpoint of realization of qubits, basic units of quantum computers (see *e.g.*[10]). The situation considered here, with preset transport currents in the banks of the single- and double-point contacts, offers the possibility to control the spontaneous bistable states in a system of $d$-wave qubits.[8]

## 2. "PARAMAGNETIC" RESPONSE OF A JOSEPHSON JUNCTION TO THE TRANSPORT SUPERCURRENT

The system, which we study, is a weak link between two superconductors with different phases of the order parameter, $\phi_1$ and $\phi_2$, and the transport supercurrent flowing parallel to the contact plane. Let us consider a weak link in the form of a small orifice in the partition between two superconductors, which is firstly subjected to the order parameter phase difference $\phi = \phi_2 - \phi_1$ (see Fig. 1a). The existence of the phase difference results in the flowing of Josephson current, which was shown to be characterized by non-sinusoidal current-phase dependence and by the suppression of the

### Vortex-like Current States in Josephson Ballistic Point Contacts

absolute value of the order parameter $\Delta = \Delta(x,y)$ in the vicinity of the orifice.[11] (The order parameter has a minimum at the orifice $\Delta = \Delta_\phi$ and a maximum far from it $\Delta = \Delta_0$; $\Delta(x,y) \in [\Delta_\phi, \Delta_0]$; $\Delta_\phi \in [0, \Delta_0]$.) The current distribution is shown at Fig. 1 (how it is calculated is explained in the next section). If there is no phase difference ($\phi = 0$), but there is homogeneous current flow in the superconductors along the boundary, the existence of the orifice does not affect the flow (see Fig. 1b). Our approach is to study the coexistence of the two factors: phase difference $\phi$ and transport supercurrent (which is parameterized by the superfluid velocity $v_s$). In this situation the local suppression of the order parameter $\Delta(x,y)$ in the vicinity of the orifice at certain relation between $\phi$ and $v_s$ (i.e. at $v_s > \Delta_\phi/p_F$, where $p_F$ is the Fermi momentum) would result in the local violation of the Landau criterion, which means the creation of unpaired quasiparticles; these quasiparticles carry current in the direction opposite to the transport supercurrent.[9] This countercurrent, being a result of the interference in a ballistic structure, is a dissipationless current. The mechanism of depairing described here is non-thermal.

A phenomenon similar to the one described here, was studied for a $d$-wave superconductor in an external magnetic field experimentally[12] and theoretically.[13] Far from the boundary the conventional diamagnetic current flows, induced by the magnetic field. But the anisotropy of the pairing potential results in the suppression of the order parameter close to the boundary and the appearance of surface quasiparticle states follows. These quasiparticle states carry a paramagnetic current counter-flowing to the diamagnetic one. The density of the paramagnetic current can substantially exceed the density of the diamagnetic one.

In our problem the transport supercurrent is analogous to the diamagnetic supercurrent in the problem of the $d$-wave superconductor in an external magnetic field; however the order parameter suppression is not due to the anisotropy of the pairing potential, but is rather due to the non-zero phase difference on the weak link. Hence the interface unpaired quasiparticle states are realized even at the contact of two conventional superconductors (in the absence of the order parameter anisotropy). In our problem the appearance of the countercurrent can be considered as the paramagnetic response of a Josephson junction to the transport supercurrent. In our previous articles[7–9] we studied the countercurrent at a weak link of both conventional and unconventional ($d$-wave) superconductors. It was shown that when the density of the countercurrent on a weak link exceeds the density of the transport supercurrent far from it, then the vortex-like formations appear in the current distribution. Here we continue this study in the geometry of the double point contact, shown in Fig. 1c.

A.N. Omelyanchouk, S.N. Shevchenko, Yu.A. Kolesnichenko

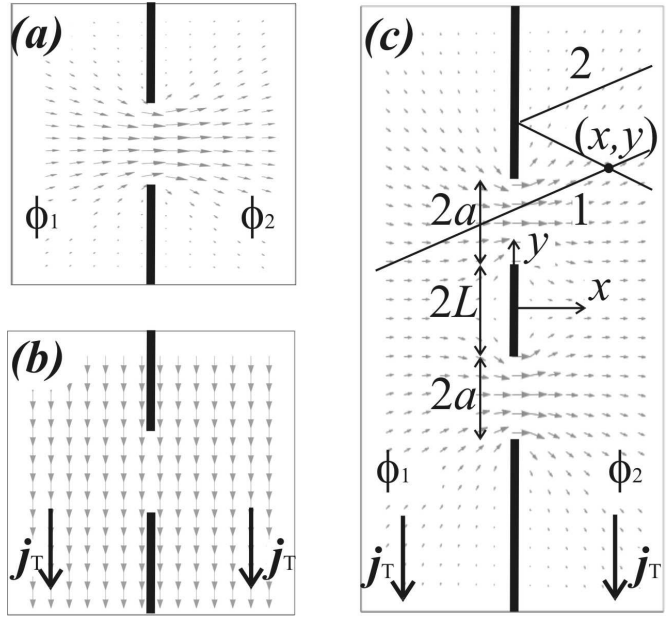

Fig. 1. Models of a single point contact (a,b) and double point contact (c) between two superconductors with different phases of the order parameter, $\phi_1$ and $\phi_2$ (a) and with homogeneous transport supercurrent $j_T(v_s)$ in the banks. The current distributions are shown when either the phase difference is present (in the absence of a transport supercurrent) (a,c) or when the transport supercurrent is present (b).

## 3. DESCRIPTION OF THE SYSTEM AND BASIC EQUATIONS

We consider two superconductors separated by an impenetrable thin partition with two identical orifices (or slits in 2D case) transparent for electrons (see Fig. 1c). The system is characterized by the ratio of the distance between orifices to their size, $L/a$: when $L = 0$, we have a single point contact (of size $4a$); when $L/a \ll 1$, then it is the model to describe a single orifice with impurity; when $L/a \sim 1$, there are two close orifices; and when $L/a \gg 1$, the orifices do not influence each other and can be considered as two single point contacts.

We describe the coherent current states in the system by solving the Eilenberger equations for quasiclassical Green functions.[14] The model of the double point contact, shown at Fig. 1c, assumes that the contact radius $a$ is smaller than the coherence length. This allows us to treat the system in the non-self-consistent approximation, which does not take into account the space dependence of the order parameter in the first approximation.

# Vortex-like Current States in Josephson Ballistic Point Contacts

The current in each point $(x,y)$ is calculated by integrating the contribution from each quasiclassical electron trajectory. We consider a clean junction, which means that each trajectory either goes trough one of the two orifices (so-called, transit trajectories, denoted by "1" in Fig. 1c) or is reflected from the partition (non-transit trajectories, "2"). In the non-self-consistent approximation transit trajectories can be described by the matrix Green function taken at the contact, $\widehat{G}_\omega(0)$, while non-transit trajectories can be described by the Green function taken far from the contact, $\widehat{G}_\omega(\infty)$. The total current density in each point in the vicinity of the weak link is given by the relations:[8,15]

$$\mathbf{j}(x,y) = 4\pi e N_0 v_F T \sum_{\omega_n > 0} \langle \hat{\mathbf{v}} Im g_\omega \rangle_{\hat{\mathbf{v}}}, \tag{1}$$

$$g_\omega(\mp\infty) = \frac{\tilde{\omega}}{\Omega_{L,R}}, \tag{2}$$

$$g_\omega(0) = \frac{\tilde{\omega}(\Omega_L + \Omega_R) - i \cdot \text{sgn}(v_x)\Delta_L\Delta_R \sin\phi}{\Omega_L\Omega_R + \tilde{\omega}^2 + \Delta_L\Delta_R \cos\phi}, \tag{3}$$

where $N_0$ is the density of states at the Fermi level, $\langle ... \rangle_{\hat{\mathbf{v}}}$ denotes the integration (averaging) over the directions of Fermi velocity $\mathbf{v}_F$, $\hat{\mathbf{v}} = \mathbf{v}_F/v_F$ is the unit vector in the direction of $\mathbf{v}_F$, $\omega_n = \pi T(2n+1)$ are Matsubara frequencies, $\tilde{\omega} = \omega_n + i\mathbf{p}_F\mathbf{v}_s$, $\Omega_{L,R} = \sqrt{\tilde{\omega}^2 + \Delta_{L,R}^2}$, and $\Delta_{L,R}$ stands for the order parameter in the left (right) bank. The current density $\mathbf{j}$ is defined by $(1,1)$-component of the matrix Green function: $G_\omega^{11} \equiv g_\omega$. Depending on the direction of the electron velocity $\hat{\mathbf{v}}$, as it is described above, the function $g_\omega$ should be taken $g_\omega = g_\omega(0)$ for transit trajectories or $g_\omega = g_\omega(\mp\infty)$ for non-transit trajectory coming from the left (right) bank.

In the case of the contact of two conventional superconductors analytical expressions can be derived for the current density at the contact plane at zero temperature,[9] which is written below for $p_F v_s < \Delta_0$:

$$j_x(0) = j_{c,0} \text{sgn}(\cos\frac{\phi}{2}) \sin\frac{\phi}{2}\left(1 - \Theta(p_F v_s - \Delta_\phi) f\left(\frac{\Delta_\phi}{p_F v_s}\right)\right), \tag{4}$$

$$j_y(0) = -\frac{2}{3}|e|N_0 v_F p_F v_s + \Theta(p_F v_s - \Delta_\phi) j_{c,0} \sin\frac{\phi}{2}\left(1 - \left(\frac{\Delta_\phi}{p_F v_s}\right)^2\right). \tag{5}$$

Here $j_{c,0} = (\pi/2)|e|N_0 v_F \Delta_0$ is the critical Josephson current at $T = 0$ and $v_s = 0$; $\Delta_\phi = \Delta_0 |\cos(\phi/2)|$ stands for the phase-dependent value of the order parameter at the contact; $f(x) = (2/\pi)\left(\arccos x - x\sqrt{1-x^2}\right) > 0$. The theta function $\Theta(p_F v_s - \Delta_\phi)$ means that at $p_F v_s < \Delta_\phi$ (i.e. at $\phi \in (\phi_1, \phi_2)$,

A.N. Omelyanchouk, S.N. Shevchenko, Yu.A. Kolesnichenko

where $\phi_1 = 2\arccos\frac{p_F v_s}{\Delta_0}$, $\phi_2 = 2\pi - \phi_1$) the Josephson current is the same as in the absence of the transport supercurrent and the current density along the contact is equal to the transport supercurrent density. At $p_F v_s > \Delta_\phi$ for a fixed value of $\phi$, the Josephson current $j_x$ is suppressed by the transport supercurrent (compared to the Josephson current in the absence of the transport supercurrent): $j_x(v_s) < j_x(v_s = 0)$, and the countercurrent $\widetilde{j}$ (the second term in Eq. 5) appears so that the total tangential current density at the contact $j_y(0)$ consists of the transport supercurrent $j_T$, carried by the condensate, and of the countercurrent $\widetilde{j}$, carried by nonthermal interface-induced quasiparticles. If there is a current at the contact which flows in the direction opposite to the direction of the current far from the contact (when $\widetilde{j} > j_T$), then the current distribution pattern contains vortex-like formations.

We note that the current density at the plane of one orifice $j_{x,y}(0)$ is not influenced by the presence of the other in the model considered here, since it is calculated from the trajectories which do not go through the second orifice. It follows that the current-phase dependence is identical for a single point contact and for double point contacts, but the current distribution pattern is not. We further note that the non-transit trajectories give non-zero input in the current $\mathbf{j}(x,y)$ when $v_s \neq 0$; this is in contrast to the case in the absence of transport supercurrent ($v_s = 0$), when $Img_\omega(\infty) = 0$ and only transit trajectories contribute to the current.[15]

Making use of Eqs. 1-3, we calculate and plot current distributions both for the cases of the contact between conventional superconductors and of the contact between $d$-wave superconductors.

## 4. THE CURRENT DISTRIBUTION PATTERNS

### 4.1. DOUBLE POINT CONTACT BETWEEN CONVENTIONAL SUPERCONDUCTORS

In the absence of the transport supercurrent $j_T(v_s)$ (i.e. at $v_s = 0$) the current distribution in the case of a single point contact is symmetric with respect to the axis of the contact and is characterized by the concentration of the current density in the vicinity of the contact (orifice), which is illustrated at Fig. 1a. Piercing of the second orifice does not influence the current distribution significantly (see Fig. 1c).

The current distribution is less trivial when the transport supercurrent is present ($v_s \neq 0$). As it is discussed in Sec. 2, at some relation between the phase difference and the value of the transport current, the countercurrent at the contact appears. Because the current at a point contact is counter-flowing to the current far from it, the current distribution contains the vortices.[7] In Fig. 2 we illustrate this for the case of a double point contact.

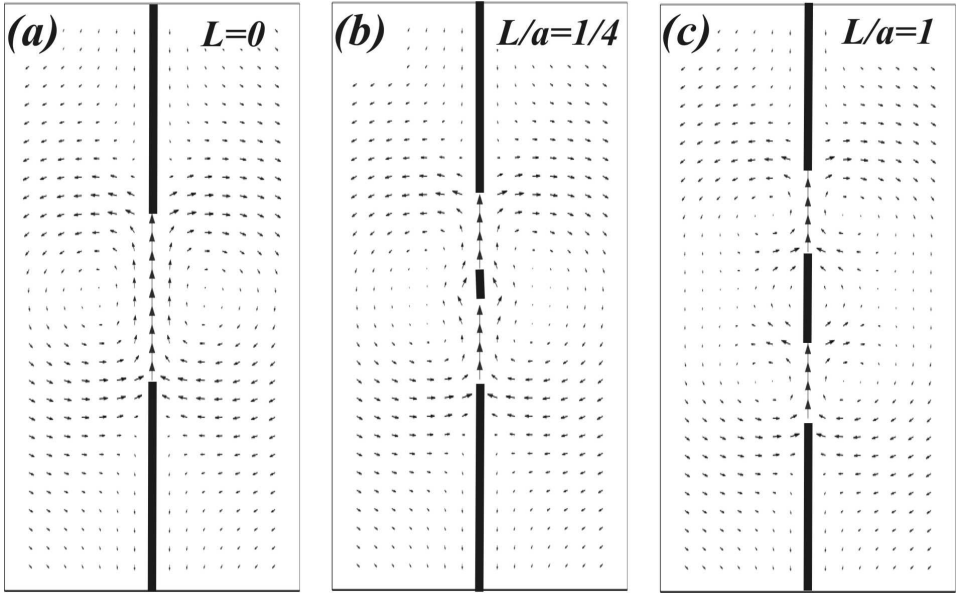

Fig. 2. Current distribution pattern in the vicinity of a double point contact between two conventional superconductors in the presence of a transport supercurrent ($v_s \neq 0$) and at the phase difference $\phi = \pi$.

### 4.2. DOUBLE POINT CONTACT BETWEEN *D*-WAVE SUPERCONDUCTORS

The current distribution for the contact of two *d*-wave superconductors can be more varied because of the anisotropy of the order parameter. Here for concreteness we consider the following orientation of the cristallographic axes: *a*-axis of the left superconductor is along the *x*-axis, and *a*-axis of the right superconductor is rotated anti-clockwise relative to it by the angle of $\pi/4$. In the absence of a transport supercurrent ($v_s = 0$) at $\phi = \pi/2$ there is no Josephson current through the junction but there is spontaneous current along the junction[16–22] and the current distribution reflects the anisotropy of the order parameter[15] (see Fig. 3a). When there is a partition between two orifices ($L \neq 0$) the anisotropy of the current distribution results in the formation of vortex-like current formations close to this partition (see Fig. 3b,c). We emphasize that this happens in the absence of a transport supercurrent, only due to the coexistence of the anisotropy and of the phase

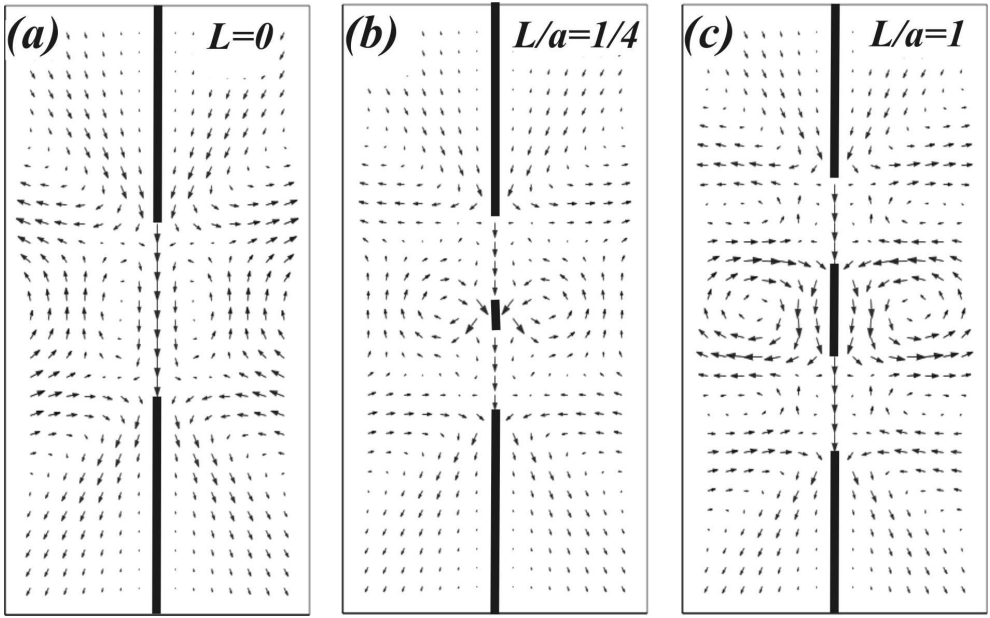

Fig. 3. Current distribution pattern in the vicinity of a double point contact between two $d$-wave superconductors in the absence of a transport supercurrent ($v_s = 0$) and at the phase difference $\phi = \pi/2$.

difference at the contact, similar to how this happens in the case of restricted geometry.[23]

In Fig. 4 we show the current distribution patterns when there is transport supercurrent in the banks of the contact ($v_s \neq 0$) at the phase difference, at which the effect of the appearance of the countercurrent is the most pronounced, at $\phi = \pi$.

### 4.3. ANALOGY WITH DC SQUID

So far we assumed that the order parameter phase difference is the same for the two slits. However the situation, when the phase difference at one junction $\phi'$ is not equal to the phase difference at another junction $\phi''$, can also be realized. This happens, when there is a magnetic flux $\Phi$, piercing the partition between the slits, as it is shown at Fig. 5a. Such a development of our model makes it analogous to the dc SQUID. At Fig. 5b,c we show the current distribution for $\phi' = -\phi'' = \pi/2$. For the contact of conventional superconductors the current circulates around the partition, which encloses the magnetic flux (Fig. 5b). The phase differences $\phi = \pm\pi/2$ in the case of two $d$-wave superconductors disoriented to each other by the angle $\pi/4$

**Vortex-like Current States in Josephson Ballistic Point Contacts**

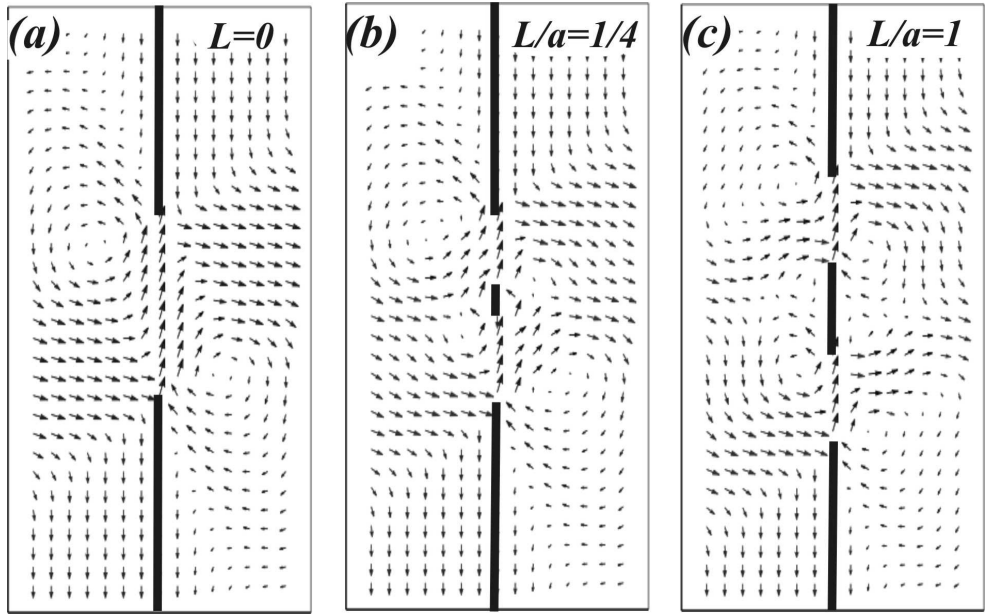

Fig. 4. Current distribution patterns in the vicinity of a double point contact between two $d$-wave superconductors in the presence of a transport supercurrent ($v_s \neq 0$) and at the phase difference $\phi = \pi$.

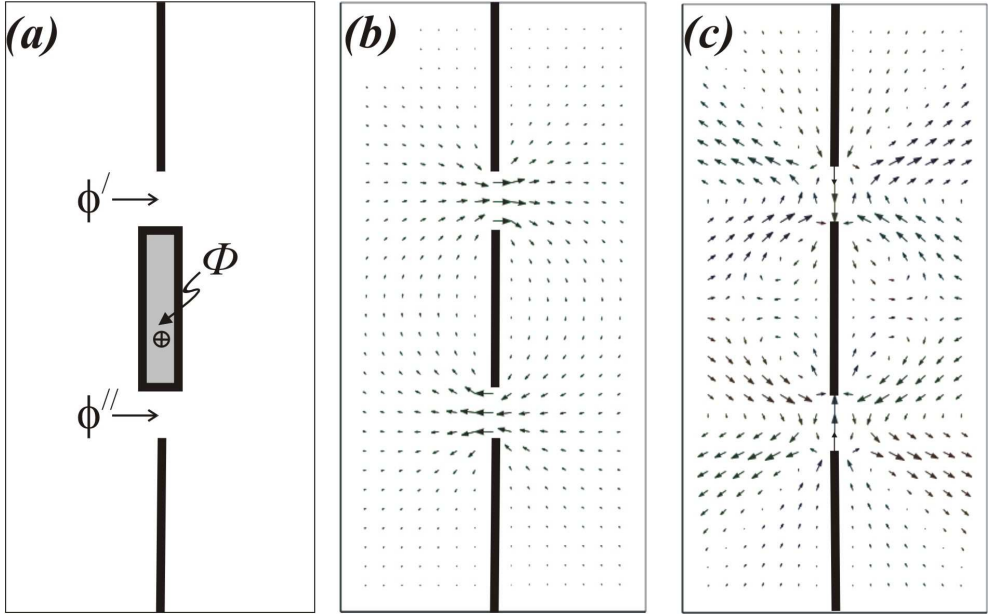

Fig. 5. Double point contact with the phase difference at one orifice $\phi'$ being not equal to the phase difference at the other orifice $\phi''$: (a) scheme, (b) and (c) current distribution pattern for the contact of conventional and $d$-wave superconductors respectively, at $\phi' = -\phi'' = \pi/2$.

correspond to the equilibrium state of the contact with no Josephson current through the contact and with spontaneous current along it; the sign "±" corresponds here to the two possible directions of the current flow.[16–22] Thus, for the contact of d-wave superconductors (at $\phi' = -\phi'' = \pi/2$) the interface currents at two junctions are in different directions. Changing the flux by the value of half-quantum-flux would invert this currents: in this way the system with interface currents can be controlled by the magnetic flux.